%% file: main.tex
\documentclass[sigconf]{acmart}

\usepackage{amsmath,amsfonts}
\usepackage{algorithmic}
\usepackage{graphicx}
\usepackage{textcomp}
\usepackage{xcolor}
\usepackage{colortbl}
\usepackage{xspace}
\usepackage{subcaption}
\usepackage{bbm}
\usepackage{fontawesome}
\usepackage{listings}
\usepackage{threeparttable}
\usepackage[ruled,linesnumbered]{algorithm2e}
\usepackage{adjustbox}   
\usepackage{pgfplots}
\usepackage{tikz}
\usetikzlibrary{shapes.geometric, arrows}
\usepackage{url}
\usepackage{hyperref}
\usepackage{multirow}
\usepackage{makecell}
\usepackage{booktabs}       
\usepackage{siunitx}    
\usepackage{array}

\newcommand{\red}[1]{\textcolor{red}{#1}}
\newcommand{\blue}[1]{\textcolor{blue}{#1}}

\definecolor{my_lightgray}{rgb}{0.9,0.9,0.9}
\definecolor{my_yellow}{rgb}{1,1,0.4}

\newcommand{\ours}{\textsc{SQLGovernor}\xspace}
\AtBeginDocument{%
  }

\setcopyright{acmlicensed}
\copyrightyear{2026}
\acmYear{2026}
\acmDOI{XXXXXXX.XXXXXXX}
\acmISBN{978-1-4503-XXXX-X/2018/06}

\begin{document}

\title[\ours: An LLM-powered SQL Toolkit for Real World Application]{\ours: An LLM-powered SQL Toolkit for Real World Application}
\author{
Jie Jiang$^{1}$,
Siqi Shen$^{2}$,
Haining Xie$^{1}$,
Yang Li$^{1}$, 
Yu Shen$^{1}$,
Danqing Huang$^{1}$,
Bo Qian$^{1}$, \\
Yinjun Wu$^{3}$,
Wentao Zhang$^{2}$, 
Bin Cui$^{3}$,
Peng Chen$^{1}$
}
\affiliation{
{{$^1$}TEG, Tencent Inc.}\\
{{$^2$}Center of Machine Learning Research, Peking University}\\
$^3$School of Computer Science, Peking University\country{}
}
\affiliation{
$^1$\{zeus, hainingxie, thomasyngli, willyushen, daisyqhuang,  leonbqian, pengchen\}@tencent.com
\\
$^2$\{shensiqi1009, wentao.zhang\}@pku.edu.cn~~~~~
$^3$\{wuyinjun, bin.cui\}@pku.edu.cn\country{}
}

\renewcommand{\shortauthors}{Jie Jiang et al.}

\begin{abstract}
SQL queries in real-world analytical environments—whether written by humans or generated automatically—often suffer from syntax errors, inefficiency, or semantic misalignment, especially in complex OLAP scenarios. 
To address these challenges, we propose \ours, an LLM-powered SQL toolkit that unifies multiple functionalities—including syntax correction, query rewriting, query modification, and consistency verification—within a structured framework enhanced by knowledge management.
\ours introduces a fragment-wise processing strategy to enable fine-grained rewriting and localized error correction, significantly reducing the cognitive load on the LLM. It further incorporates a hybrid self-learning mechanism guided by expert feedback, allowing the system to continuously improve through DBMS output analysis and rule validation.
Experiments on benchmarks such as BIRD and BIRD-CRITIC, as well as industrial datasets, show that \ours consistently boosts the performance of base models by up to 10\%, while minimizing reliance on manual expertise. Deployed in production environments, \ours demonstrates strong practical utility and effective performance.
\end{abstract}



\maketitle

\input{introduction}
\input{related_work}
\input{method}
\input{experiment}
\input{conclusions}


\bibliographystyle{ACM-Reference-Format}
\bibliography{ref}


\end{document}

%% file: introduction.tex
\section{Introduction}

In real-world analytical applications, Structured Query Language (SQL) remains the primary interface for interacting with relational databases. Despite its maturity and widespread adoption, crafting accurate, efficient, and semantically aligned SQL queries—especially in complex analytical (OLAP) scenarios—remains a challenging task for both novice and experienced users alike.

OLAP workloads are central to modern business intelligence, reporting, and decision-making systems~\cite{codd1993OLAP, thomsen2002OLAP}. Even minor inefficiencies or ambiguities can lead to significant performance degradation, incorrect insights, or increased development overhead~\cite{vassiliadis2000OLAP, pedersen2001OLAP}.
SQL queries in OLAP settings typically exhibit three key characteristics. First, they perform multi-dimensional analysis using advanced operations such as roll-up and drill-down~\cite{ceci2015SQLOLAP}, resulting in highly structured and deeply nested query forms. Second, these queries operate on large-scale enterprise data~\cite{chen2012BIbigdata}, which increases computational costs and run-time unpredictability. Third, many OLAP queries are executed repeatedly—such as daily or weekly reports—making even small inefficiencies costly over time~\cite{zhan2019analyticdb}.

The repetitive and high-stakes nature of OLAP queries amplifies the need for robust, automated, and adaptive SQL post-processing solutions. Given these challenges, many tools have been developed to assist users in crafting better SQL queries, including syntax correction, query rewriting, and semantic refinement~\cite{cen2024sqlfixagent, self-debug, liu2024genrewrite, li2024llm-r2}. While large language models (LLMs) have shown great promise in translating natural language questions into SQL, their applicability across the broader spectrum of SQL-related tasks remains underexplored. Moreover, in industrial practice, many users lack deep database expertise and often produce poorly written queries that are inefficient or semantically inaccurate, further exacerbating system performance and reliability issues~\cite{banisharif2022automaticBIchatbot}.

\begin{table}[htbp!]
\centering
\vspace{-2.5mm}
\caption{Comparison among the latest SQL tools from multiple aspects. \red{$\checkmark$} indicates implemented and \blue{$\times$} indicates not implemented. Function notions-1: Fix Syntactic Error; 2. Refine Semantic Error; 3. Rewrite Query; 4. Verify SQL Pair Equivalence}
\vspace{-2.5mm}
\begin{adjustbox}{width=0.95\columnwidth}
    {
    \begin{tabular}{c|c|c|c|c}
         \toprule
         \textbf{Tools} & \makecell[c]{\textbf{Supported} \\ \textbf{Functions}} & \makecell[c]{\textbf{Knowledge} \\ \textbf{Management}} & \textbf{Scenario} & \textbf{Cost} \\ 
         \hline
         \textsc{Self-Debugging}~\cite{self-debug} & 2 & \blue{$\times$} & General & High \\
         SQLFixAgent~\cite{cen2024sqlfixagent}  & 1,2 & \red{\checkmark} & NL2SQL & Medium \\
         \textsc{Cyclesql}~\cite{cyclesql} & 2 & \blue{$\times$} & NL2SQL & Medium \\
         WeTune~\cite{wang2022wetune} & 3 & \blue{$\times$} & OLTP & Low \\
         LLM-R$^2$~\cite{li2024llm-r2} & 3 & \blue{$\times$} & General & Medium \\
         GenRewrite~\cite{liu2024genrewrite} & 3,4 & \blue{$\times$} & OLAP & Medium \\
         LimeQO~\cite{yi2025limeQO} & 3 & \blue{$\times$} & Offline QO & High \\
         FuncEvalGMN~\cite{FuncEvalGMN} & 4 & \blue{$\times$} & NL2SQL & Medium \\
         \hline
         \ours & 1,2,3,4 & \red{\checkmark} & \makecell[c]{General \\ Excel in OLAP} & Medium \\
         \bottomrule
    \end{tabular}}
\end{adjustbox}
\vspace{-2.5mm}
\label{tab:c1_compare}
\end{table}

\textit{C1: Productivity bottleneck caused by a fragmented ecosystem.}
As summarized in Table~\ref{tab:c1_compare}, existing SQL tools typically offer isolated functionalities and lack a unified framework that addresses the full spectrum of SQL data governance tasks—ranging from syntax correction and semantic refinement to query rewriting and consistency verification.
For instance, recent works such as SQLFixAgent~\cite{cen2024sqlfixagent}, \textsc{Self-Debugging}~\cite{self-debug}, and LLM-R$^2$~\cite{li2024llm-r2} are designed with narrow scopes, focusing primarily on specific subtasks like error correction or rule-based rewriting. 
Moreover, few of these systems incorporate a structured knowledge management module to support intricate or domain-specific tasks.
This fragmented ecosystem significantly raises the barrier to entry—particularly for non-expert users. Even for experienced database practitioners, empirical studies show that managing disconnected tool-chains leads to a 30–40\% increase in manual effort due to frequent context switching, compatibility issues, and redundant operations~\cite{stackoverflow}.

\textit{C2: Lack of advanced query processing techniques tailored for OLAP.} 
Existing SQL tools mainly target general-purpose scenarios~\cite{self-debug} or other domains like OLTP~\cite{wang2022wetune}, NL2SQL~\cite{cen2024sqlfixagent, cyclesql}, and offline query optimization~\cite{yi2025limeQO}. SQL queries in different scenarios have distinct characteristics and thus require different tool capabilities. For example, OLTP queries are simple and execute within milliseconds, so the matched query rewriting methods must be extremely low-overhead.
In contrast, OLAP queries involve complex queries operating on large data volumes with longer run-times, often ranging from seconds to minutes~\cite{zhan2019analyticdb}. The query rewriting tools for OLAP workloads must carefully balance effectiveness and computational cost: lightweight algorithms~\cite{wang2022wetune, li2024llm-r2} may fail to fully capture the complexity of OLAP queries, while computationally intensive ones~\cite{yi2025limeQO} risk increasing end-to-end execution time, thus failing to meet online response requirements. This trade-off similarly affects other SQL-related tasks. Such challenges underscore the need for dedicated, scalable query processing techniques specifically tailored to OLAP.

\textit{C3: High operational cost in expert-centric knowledge lifecycle.}
The query correction and rewriting tasks require not only proficiency in SQL syntax but also deep domain knowledge and understanding of underlying database internals. Unlike simple pattern-based fixes, they involve context-aware reasoning and complex logic that are difficult to capture with conventional data-driven methods. Consequently, enterprises heavily depend on expert teams, incurring labor cost increases of 25\%-35\% as reported by Gartner~\cite{gartner_2022}. Furthermore, maintaining and updating expert knowledge bases to keep pace with evolving database technologies and query patterns introduces additional overhead, raising the adoption threshold for ordinary users and limiting scalability. The scarcity of experts and the time-consuming nature of knowledge curation exacerbate these challenges, making sustainable and efficient SQL governance elusive for many organizations.

In summary, the current landscape lacks a comprehensive and practical SQL toolkit, which utilize evolving knowledge with fewer human efforts.

To address \textit{C1}, we propose an LLM-powered SQL toolkit that unifies multiple functionalities within a structured framework enhanced by knowledge management. 
Users can either select individual tools for specific tasks or use an end-to-end pipeline that orchestrates multiple tools in a coordinated, use-case-driven manner.
By consolidating diverse functionalities into a single platform, our approach eliminates the fragmentation in existing SQL tool-chains, significantly reducing deployment overhead, manual effort, and the barrier to entry.

To address \textit{C2}, we adopt a dual approach that applies validated rules as guidance when appropriate, while permitting the LLM autonomous operation otherwise. Additionally, for particularly long and structurally complex queries, we propose a ``fragment processing'' strategy to reduce the chance of LLM hallucinations and lower the cost of using the LLM.

To address \textit{C3}, we propose an expert-guided iterative self-learning mechanism to maintain a dynamic knowledge base for SQL tasks. The LLM agent analyzes DBMS outputs to generate new rules, identifying unseen error types from failed SQL and discovering rewriting strategies for inefficient queries. These rules are periodically verified by experts and integrated into the knowledge base for continuous improvement.

The main contributions of this paper can be summarized as follows:
\begin{enumerate}
    \item \underline{Unified Framework}: To the best of our knowledge, \ours is the first comprehensive LLM-based SQL toolkit with a knowledge management module. It provides four core functionalities powered by a hybrid self-learning mechanism, thereby improving both user productivity and SQL quality.
    \item \underline{Fragment Processing}: We propose a fragment-wise processing strategy to address the complexity and length of OLAP queries. By localizing error detection and rewriting within individual fragments, including subqueries and CTEs, our approach enhances precision and reduces the cognitive burden on LLMs.
    \item \underline{Hybrid Self-Learning}: We introduce an expert-guided hybrid self-learning framework that enables \ours to automatically extract common pattern from execution outputs, generate and validate new knowledge with minimal expert intervention, leading to continuous performance improvement.
    \item \underline{Proven Effectiveness}: Extensive experiments on academic benchmarks and real-world industrial datasets demonstrate that \ours consistently improves the performance of base models by up to 10\% in key metrics. Deployed in production environments, \ours indicates strong practical utility and effective performance.
\end{enumerate}

%% file: related_work.tex
\begin{figure*}[htbp!]
  \centering
  \includegraphics[width=0.95\linewidth]{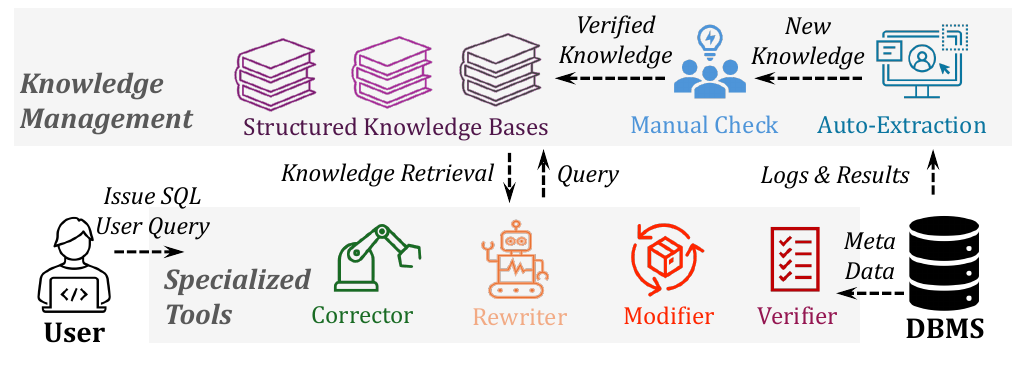}
  \vspace{-2.5mm}
  \caption{\ours integrates four specialized SQL tools and a knowledge management module into a unified framework.}
  \label{fig:overview}
  \vspace{-2.5mm}
\end{figure*}

\section{Related Work}

\subsection{Query Rewriting}

Query rewriting can be classified according to the stages of the SQL query lifecycle. A typical query goes through several phases: parsing for syntax validation, binding for schema resolution, optimization using a cost model, and execution by the database engine. Based on this lifecycle, rewriting can be applied at different levels: (1) raw SQL queries before parsing, (2) logical plans generated by the binder, or (3) physical plans produced by the optimizer within the DBMS.

Physical plan rewriting focuses on selecting the most efficient execution plan among functionally equivalent alternatives. The DBMS optimizer employs search algorithms—such as dynamic programming~\cite{graefe1995cascades, graefe1993volcano}—to explore the space of physical plans without exhaustive enumeration. A cost model is used to estimate the execution cost of each candidate plan. Recent approaches have introduced machine learning and deep learning techniques~\cite{kipf2018learned, trummer2021skinnerdb, sun2019end} to enhance cost estimation and plan selection. However, these methods often suffer from long training times, limited adaptability across workloads, and high integration costs.

An alternative approach that preserves the existing optimizer architecture is Bao~\cite{marcus2021bao}, which integrates a Tree Convolutional Neural Network~\cite{mou2016convolutional} with Thompson sampling to guide the selection of better hint sets. This hybrid approach improves plan generation without requiring a complete redesign of the optimizer, thereby balancing innovation with system compatibility.

Logical plan rewriting involves transforming a tree-structured representation of query operations, focusing on what to compute rather than how to compute it. This process is primarily driven by rule-based pattern matching, where heuristic rules—such as predicate push-down and operation merging—guide the transformation~\cite{levy1994query, pirahesh1992extensible, muralikrishna1992improved}. Recent efforts aim to automate rule discovery. WeTune~\cite{wang2022wetune} uses brute-force enumeration to identify and validate new rules, enhancing the internal optimizer's capabilities. QueryBooster~\cite{bai2023querybooster} introduces a connector for user-defined rules, enabling task-specific rewriting strategies. Traditional rule application orders are often fixed and suboptimal. LR~\cite{zhou2021learnedrewrite} applies Monte Carlo Tree Search to explore effective rewriting sequences, while LLM-R$^2$~\cite{li2024llm-r2} leverages large language models to recommend context-aware rewriting rules, improving adaptability and generalization.

End-to-end SQL rewriting aims to enhance transparency and usability by rewriting queries before they enter the DBMS pipeline. This approach enables holistic transformations and avoids the limitations of local rewriting efforts. Recent studies have explored the use of Large Language Models (LLMs) to facilitate this process. The DB-GPT framework~\cite{zhou2024db-gpt} categorizes such approaches into three paradigms: in-context learning, LLM fine-tuning, and DB-specific pre-training. GenRewrite~\cite{liu2024genrewrite}, a representative in-context learning method, designs prompts to guide LLMs in SQL rewriting and stores generated natural language rules in the NLR2s repository for reuse. To mitigate LLM hallucinations, GenRewrite includes a validation and correction step to ensure the reliability of the rewritten queries.

Additionally, middleware-based rewriting has been explored to offload optimization tasks from the DBMS. This approach provides a flexible layer between the application and the database, enabling query transformation before execution~\cite{bai2023improving}. Similarly, query rewriting has been integrated into human-in-the-loop systems to support interactive data exploration, where users can iteratively refine queries based on intermediate results.

\subsection{Query Error Detection and Correction}

SQL query errors fall into two categories: syntax errors and semantic errors. Syntax errors occur when a query violates SQL's syntactical rules, preventing execution. Traditional debugging methods, as noted by Gathani et al. \cite{gathani2020debugging_survey}, lack automated correction and instead help users identify errors through techniques like visualizing intermediate results. Semantic errors arise when a query fails to return expected data, indicating a mismatch between the query's output and the user's intent. Verifying query equivalence is crucial in NL2SQL conversion \cite{pourreza2024chase_SQL, talaei2024chess_SQL, gao2024xiyan_SQL} and query rewriting models \cite{slabcity, liu2024genrewrite, wang2022wetune}. Existing SQL equivalence provers use algebraic representations to verify query equivalence by solving mathematical problems~\cite{ding2023proving, zhou2019automated}, offering high reliability but at high computational cost. For loosely bounded verification, heuristic rules and counterexample construction are employed~\cite{slabcity}, while some studies leverage LLMs for reasoning and judgment~\cite{liu2024genrewrite}.

%% file: method.tex
\section{Framework Design}

As illustrated in Figure~\ref{fig:overview}, the architecture of \ours comprises four specialized tools, each tailored to address a specific category of SQL-related tasks. These tools are supported by a knowledge management module that analyzes historical data and extracts actionable insights to guide error correction, query rewriting, and semantic refinement.

Specifically, we focus on three common issues encountered during SQL execution in real-world applications:
\begin{itemize}
\item Resolving execution failures caused by syntax errors,
\item Rewriting inefficient queries that result in excessively long execution times,
\item Modifying SQL to better align with user intent.
\end{itemize}

To systematically address these challenges, we design the following core components:
\begin{itemize}
    \item \textbf{Query Rewriter}: Transforms a given SQL query into a semantically equivalent but more execution-efficient version.
    \item \textbf{Equivalence Verifier}: Validates whether two SQL queries are semantically equivalent—i.e., produce the same results regardless of database content.
    \item \textbf{Query Modifier}: Enables users to express desired modifications through a natural language interface.
    \item \textbf{Syntax Error Corrector}: Detects and repairs syntax-level errors in SQL queries.
\end{itemize}

\ours operates as follows: Users identify the issue affecting an SQL query and select the appropriate tool from the toolkit. The selected tool then processes the query, optionally consulting the knowledge management module to enhance accuracy and efficiency. Once processed, the revised SQL is returned to the user for re-execution on the DBMS.

The execution outputs—including query results, success status, error messages, and performance metrics—are collected and fed back into the knowledge base. This closed-loop mechanism supports continuous learning and adaptation, enabling the system to evolve alongside changing workloads and environments. As a result, \ours maintains high robustness and effectiveness over time.

\section{Knowledge Management}

To enhance the effectiveness of LLMs in SQL-related tasks, we design a knowledge management module paired with dedicated recall and retrieving techniques. 
\ours benefits from past experiences and domain-specific insights to reduce redundant computations and improve both accuracy and consistency. To further automate knowledge acquisition and minimize reliance on manual expert input, we implement structured knowledge bases inspired by the self-reinforcing data flywheel mechanism~\cite{data_flywheel}. Each knowledge base continuously accumulates and updates rules from both successful and failure cases, enabling the system to improve over time with minimal supervision.

\begin{figure}[htbp!]
    \centering
    \begin{subfigure}[b]{\linewidth}
        \centering
        \includegraphics[width=0.9\linewidth]{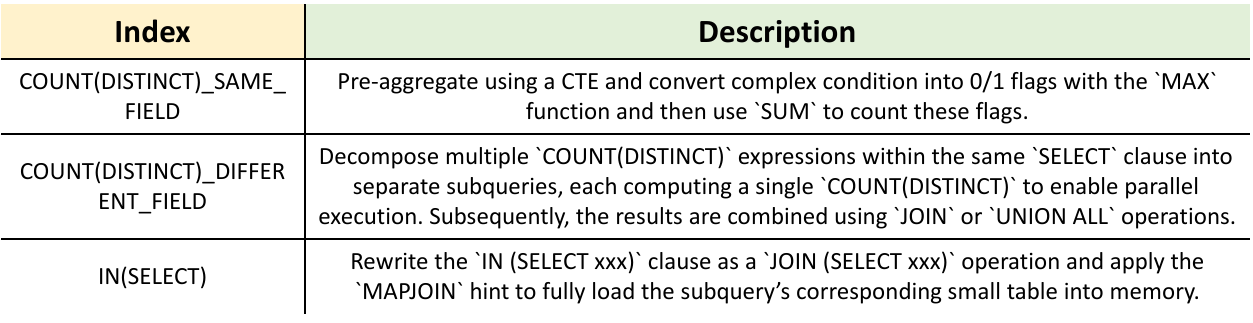}
        \vspace{-1.5mm}
        \caption{``Rules''}
        \label{subfig:rule_example}
    \end{subfigure}
    \begin{subfigure}[b]{\linewidth}
        \centering
        \includegraphics[width=0.9\linewidth]{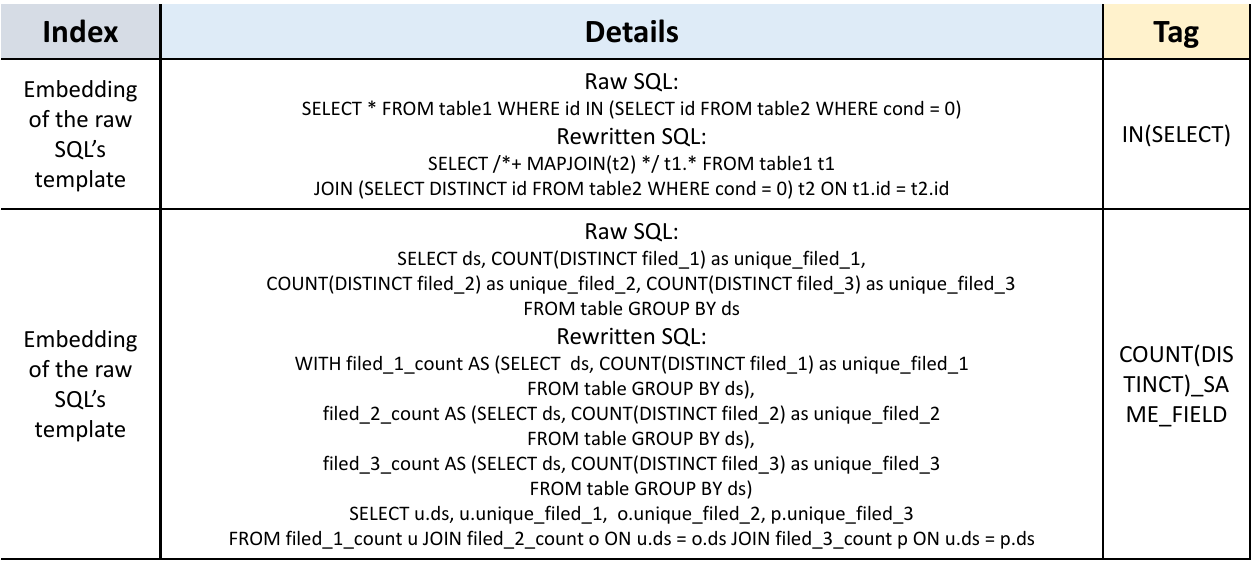}
        \vspace{-1.5mm}
        \caption{``Historical Data''}
        \label{subfig:data_pipeline}
    \end{subfigure}
    \vspace{-6.5mm}
    \caption{Demonstration of the knowledge base serving the Query Rewriter tool.}
    \label{kb_example}
    \vspace{-2.5mm}
\end{figure}

\subsection{Structure of the Knowledge Base}

The knowledge base is carefully structured to meet the diverse needs of different SQL tools, with each tool having its own dedicated repository. Each repository is divided into two sub-modules: ``Rules'' and ``Historical Data''. The ``Rules'' sub-module contains structured entries that provide actionable guidance for the LLM in addressing specific SQL-related tasks. These rules include transformation patterns, rewriting strategies, and syntactic corrections, enabling the model to apply well-defined solutions in a consistent manner. The ``Historical Data'' sub-module stores high-quality, representative cases collected from real-world applications of the SQL tools. By analyzing these past examples, the LLM can recognize recurring patterns and adopt strategies that have proven effective in similar contexts.

Figure~\ref{kb_example} illustrates the structure of the knowledge base used by the Query Rewriter. Each rule entry consists of two fields: \textit{index}, used for efficient retrieval, and \textit{description}, which provides a detailed explanation of the rule’s application and scope. Each historical data entry includes three fields: \textit{index} (for retrieval), \textit{details} (a full description of the case), and \textit{tag}, which links the case to relevant rule indices, thereby facilitating cross-referencing between observed problems and applicable solutions.

\subsection{Knowledge Storage and Retrieval}

To effectively support the diverse types of knowledge entries, we design tailored storage and retrieval strategies for each sub-module of the knowledge base.
For the ``Historical Data'' module, all entries are stored in a vector database—specifically StarRocks~\cite{starrocks}. Retrieval is based on the structural and semantic similarity between SQL templates. To facilitate this, we first extract the template of each SQL query by replacing concrete identifiers (e.g., table and column names) with symbolic placeholders. This abstraction allows the encoder to focus on high-level patterns rather than surface-level variations. During retrieval, the input SQL is similarly transformed into a template, encoded into a vector representation, and used to retrieve the top-$k$ most similar historical cases via cosine similarity. These candidates are further filtered using the \textit{Tag} field to ensure alignment with the specific task or error type. 

In contrast, the ``Rules'' module employs different strategies depending on its application context. For the Query Rewriter, each rule is associated with a unique label (e.g., \texttt{IN(SELECT)}) that captures its applicability condition. We store these rules in an ElasticSearch~\cite{elasticsearch2018elasticsearch} database to enable efficient exact matching during retrieval. 
For the Syntax Error Corrector, exception categories (e.g., \texttt{RuntimeException}, \texttt{SqlValidatorException}) are often too coarse-grained to be informative. Therefore, we retain more detailed error messages (e.g., \texttt{``SqlValidatorException: INNER, LEFT, RIGHT or FULL join requires a condition (NATURAL keyword or ON or USING clause)''}) as retrieval targets. These messages are stored in a vector database. Given the verbosity of real-world SQL execution logs, we first apply regular expressions to extract key information before encoding it into vector representations for retrieval.

\subsection{Hybrid Self-Learning Mechanism}

To ensure the knowledge base is both effective and actionable, we adopt a multi-source initialization strategy tailored to each sub-module.
For the ``Rules'' sub-module, initialization involves aggregating knowledge from diverse sources and organizing it into structured entries. For the Query Rewriter, the rule set is primarily sourced from domain experts and is designed to complement the built-in optimization rules of the DBMS. These expert-defined rules have been rigorously validated to ensure their practical effectiveness in real-world OLAP scenarios.
For the Syntax Error Corrector, the rule base is initialized using frequently asked questions (FAQs) and common error-handling guidelines from technical documentation. These are formalized into <exception, fixing action> pairs and further mapped to the <index, description> structure for retrieval compatibility.
For the ``Historical Data'' sub-module, initialization is conducted in two ways: (1) manually curating high-quality cases that align with existing rules, and (2) extracting representative queries from execution logs. These cases are selected based on criteria such as execution frequency, complexity, and historical performance, ensuring their relevance and utility in future inference tasks.

Furthermore, both sub-modules support incremental updates through the self-learning mechanism, allowing the system to refine its knowledge over time based on new data and user feedback. 
Specifically, we design a rule update paradigm that automatically extracts supplementary rules by analyzing SQL queries that fail to meet user requirements—such as those resulting in execution errors, returning incorrect results, or exhibiting excessive runtime. A heuristic pattern recognition-based data filtering algorithm is first applied to extract relevant features (e.g., error messages, query structures, execution times) from execution logs. An LLM agent then analyzes this information, identifies common inefficiencies or mistakes, and generates structured knowledge entries. The prompt used for rule generation is as follows:
\begin{lstlisting}[caption={Prompts for generating rules.}]
Task Description: 
    You are provided with an SQL query along with its execution outputs (e.g. logs and results) from DBMS. Your task is to analyze the logs and results to identify potential errors or inefficiencies in the query.
  -
Instruction:
    1. Review the execution logs and results to determine whether the query contains errors or inefficiencies.
    2. For each confirmed problem, try to distill it into a generalizable rule, including the abstract problem pattern, detailed description, and its solution.
    3. Convert your findings into JSON format, where the key is the problem pattern used as index and the value is a detailed description of the problem along with possible corrective actions.
  -
Demonstration:
    {Few-shot examples curated from previously validated rule entries.}
  -
Question:
    {SQL query and user query if necessary.}
  -
Execution Outputs:
    {From logs and results.}
\end{lstlisting}

To ensure accuracy and reliability, experts verify the newly generated rules based on predefined conditions. We implement a threshold-triggered mechanism to maintain the relevance of the knowledge base. Verification is triggered when either: 
(a) the number of new rules exceeds a threshold $t_1$, or 
(b) the time elapsed since the last update surpasses a threshold $t_2$, where:
\begin{align}
  t_1 &= \lfloor \lambda \cdot \sqrt{N_{\text{current}}} \rfloor \\
  t_2 &= \beta \cdot \mathbb{E}[\Delta t_{\text{historical}}]
\end{align}
with $\lambda=2.5$ controlling capacity scaling and $\beta=1.3$ for temporal adaptation. Here $N_{\text{current}}$ denotes the current number of rules, and $\mathbb{E}[\Delta t_{\text{historical}}]$ represents the expected historical update interval.
To avoid redundancy, semantically equivalent rules are identified and clustered through the following process: 
(a) The description field of rule $r_i$ is encoded using RoBERTa-base~\cite{liu2019roberta}.
(b) Compute the pairwise similarity.
(c) Rules are grouped around $r_i$ into $C_k(i)$ based on the similarity score using DBSCAN~\cite{schubert2017dbscan}.
(d) Merge semantically equivalent rules using centroid synthesis.
\begin{gather}
    \mathbf{e}_i = \text{RoBERTa}(r_i.\text{get(description)}) \in \mathbb{R}^{768} \\
    s(r_i, r_j) = \frac{\mathbf{e}_i^\top \mathbf{e}_j}{\|\mathbf{e}_i\| \|\mathbf{e}_j\|} \\
    r_{\text{new}} = \arg\min_{r \in C_k(i)} \sum_{r_i \in C_k(i)} \| \mathbf{e}_r - \mathbf{e}_{r_i} \|_2
\end{gather}
The updated rules are then applied to subsequent tasks, generating new data that perpetuates the improvement cycle. 

The update mechanism for the ``Historical Data'' sub-module is relatively straightforward. As mentioned earlier, for rules that have been validated by experts, we incorporate the corresponding instances into the historical data. This continuous cycle of knowledge collection, validation, and application ensures that the knowledge base remains up-to-date and effective, thereby enhancing the overall performance of the LLM-based SQL toolkit.

\section{Specialized SQL Tools}

This section introduces our core design principle—the fragment processing strategy—which enables fine-grained, modular analysis for complex SQL tasks. Built upon this foundation, we further present four specialized tools in \ours, each targeting a key subtask. The following subsections provide detailed descriptions of these tools and their underlying mechanisms.

\subsection{Fragment Processing Strategy}

Modern SQL queries, especially in OLAP workloads, often exhibit deeply nested structures with multiple layers of subqueries and Common Table Expressions (CTEs). To enable scalable and precise analysis, we propose a recursive fragment processing strategy that decomposes an input SQL query into smaller, self-contained fragments. Each fragment is analyzed independently under the same procedure, significantly reducing the reasoning complexity for LLM-based agents.

Concretely, the strategy operates as follows: the input query is first partitioned into a main query and a collection of CTEs. Each of these components is then recursively parsed to extract subqueries, which are likewise treated as fragments. For every fragment, the system performs a localized analysis and stores the result as a tuple of the form \texttt{<fragment\_id, analysis\_res>}.

The pseudocode in Algorithm~\ref{alg:fragment-processing} outlines the core idea for fragment processing.

\begin{algorithm}[htbp!]
\caption{Fragment Processing Strategy}
\label{alg:fragment-processing}
\KwIn{SQL query $Q$}
\KwOut{Analysis result set $S$}

$S \gets \emptyset$ \tcp*{Initialize result set}

$D \gets \emptyset$

$Q_{main}, \{Q_{cte_j}\} \gets \texttt{divide\_CTE}(Q)$

\If{$Q_{main} = \emptyset$}{
    \Return $\emptyset$
}

$\{Q_{sub_i}\} \gets \texttt{parse\_subqueries}(Q_{main})$

\ForEach{$Q_{sub} \in \{Q_{sub_i}\}$}{
    $S_{sub} \gets \texttt{fragment\_processing}(Q_{sub})$ \\
    $S \gets S \cup S_{sub}$
}

\ForEach{$Q_{cte} \in \{Q_{cte_j}\}$}{
    $S_{cte} \gets \texttt{fragment\_processing}(Q_{cte})$ \\
    $S \gets S \cup S_{cte}$
}

\tcp{Main query analysis (details omitted)}

\Return $S$
\end{algorithm}

\subsection{Query Rewriter}

The Query Rewriter tool performs query optimization via a two-stage process: evaluation and rewriting.
In the evaluation stage, the tool analyzes the input query to determine whether it exhibits inefficiencies. Specifically, it localizes the bottlenecks to specific fragments and proposes targeted rewriting strategies. 
In the rewriting stage, the tool applies the selected rewriting strategies to generate an optimized version of the original query. By leveraging the fragment-level structure produced during the fragment processing stage, the rewriting is both context-aware and fine-grained, ensuring that improvements are applied precisely without altering the query semantics.

\subsubsection{Evaluation}

\begin{figure}[htbp!]
    \centering
    \includegraphics[width=0.95\linewidth]{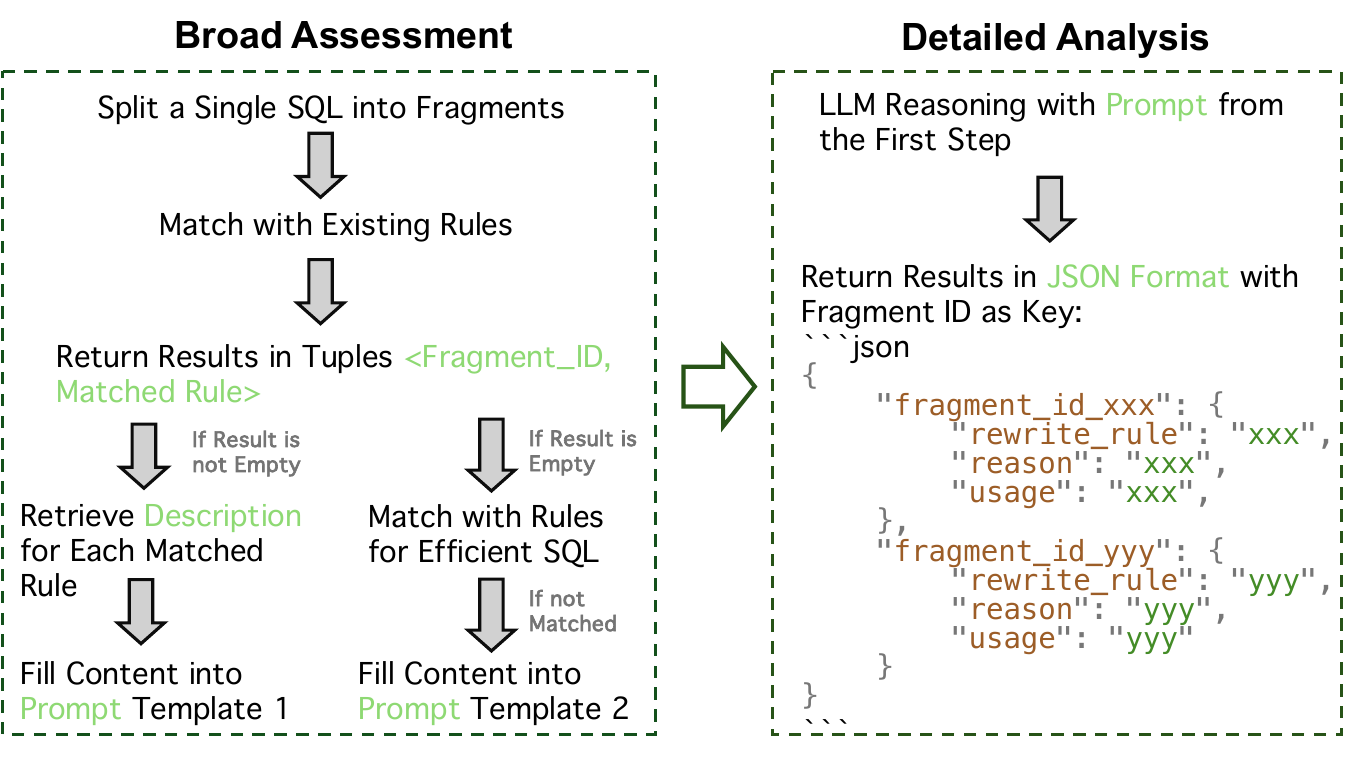}
    \vspace{-2.5mm}
    \caption{Workflow of the two-stage evaluation process.}
    \vspace{-2.5mm}
  \label{evaluate_efficiency}
\end{figure}

As illustrated in Figure~\ref{evaluate_efficiency}, the evaluation process combines rule-based pattern matching with LLM-driven reasoning to deliver both precise and innovative optimization strategies—enhancing the overall efficiency of complex SQL queries.

If any rules match during the initial phase, their detailed descriptions are retrieved from the knowledge base and combined with the corresponding fragments into a structured prompt template (referred to as Scenario 1). If no rules are matched, the system evaluates the query against a set of ``already efficient SQL'' rules. A successful match indicates that the query is already optimal and does not require further rewriting efforts; otherwise, it suggests potential optimization opportunities that may not be covered by existing rules. In such cases, the query is passed to another predefined prompt template (referred to as Scenario 2).

Following the broad assessment, a detailed analysis is conducted using the LLM to refine and expand upon the rewriting suggestions. Specifically: 
(1) For Scenario 1, the prompt instructs the LLM to evaluate the applicability of each suggested rule, providing justifications and transforming general recommendations into actionable instructions where appropriate.
(2) For Scenario 2 , the prompt directs the LLM to analyze the intent of the original query and explore alternative, more efficient formulations that preserve semantic equivalence.
The output of this stage is standardized in \texttt{JSON} format, enabling seamless integration with the subsequent rewriting module.

To clarify the ``fragment processing'' design, Listing~\ref{motivation_example} provides a representative example. This SQL query contains six subqueries across different levels, with the deepest level of nesting being four. We have numbered all subqueries according to the order in which they are analyzed. The analysis results are as follows: Fragments 1-3 did not match any rules and meet the criteria for efficient SQL. Fragment 4 matched the \texttt{SAME\_TABLE\_JOIN} rule, as it was detected to scan the same table (e.g., \texttt{tb0}) twice. Fragments 5-6 also did not match any rules and are considered efficient. The outermost query (i.e., fragment 7) satisfies a rule involving a \texttt{LEFT JOIN} and an \texttt{IS NOT NULL} condition. These two defects were further confirmed by the LLM.
\begin{lstlisting}[
  language=SQL, 
  commentstyle=\color{gray},
  keywordstyle=\color{blue}\fontsize{7}{7}\ttfamily,
  basicstyle=\fontsize{7}{7}\ttfamily,
  stringstyle=\color{red},
  frame=single,
  caption={Show the working mechanism of the fragment processing design.},
  label={motivation_example},
  morekeywords={IFNULL},
  numbers=left,          
  numberstyle=\tiny\color{gray},  
  stepnumber=1,
  numbersep=5pt]
-- Fragment 7, Line 2-27
SELECT tb0.c0, 
-- Fragment 5, Line 4
(SELECT tb3.c1 - tb3.c2 FROM tb3 WHERE tb3.ds = tb1.ds),
-- Fragment 6, Line 6
(SELECT AVG(tb4.c3) FROM tb4 WHERE tb4.ds = tb1.ds AND tb4.c3 > 100)
FROM tb1 
LEFT JOIN tb2 ON tb1.ds = tb2.ds 
WHERE tb2.ds is NOT NULL AND
tb1.dcrs <= 
(
  -- Fragment 4, Line 13-27
  SELECT IFNULL(t1.c1 / t2.c2, 100) AS dcrs
FROM
  -- Fragment 2, Line 16-22
 (SELECT MIN(c) AS c1
  FROM
    -- Fragment 1, Line 19-22
    (SELECT COUNT(*) AS c, ds
     FROM tb0
     WHERE ds >= '1014' AND ds < '1016'
     GROUP BY ds)) AS t1,
  -- Fragment 3, Line 24-26
 (SELECT COUNT(*) AS c2
  FROM tb0
  WHERE ds = '1016') AS t2
)
\end{lstlisting}

\subsubsection{Rewriting}

As previously described, the Query Rewriter identifies potential inefficiencies in the input SQL and generates a set of actionable rewriting suggestions during the evaluation phase. In addition, the system retrieves relevant historical rewriting examples from the knowledge base by matching the current SQL fragments and their associated optimization rules. 
The LLM then integrates this information and synthesizes into a semantically equivalent yet execution-efficient SQL query that incorporates the suggested optimizations.

Listing~\ref{rewriting_example_listing} illustrates the rewritten SQL for the query presented in Section 5.2.1. Specifically, the pattern involving a \texttt{LEFT JOIN} combined with the \texttt{IS NOT NULL} condition is replaced with an \texttt{INNER JOIN}. This transformation is effective because an \texttt{INNER JOIN} naturally filters out rows with null values, thus achieving the same result as the original query but with a more efficient join operation.
Furthermore, the two separate scans of table \texttt{tb0} are merged into a single scan to reduce I/O overhead. The \texttt{WHERE} and \texttt{SELECT} clauses are appropriately adjusted to preserve the query's semantic correctness.

\begin{lstlisting}[
  language=SQL, 
  commentstyle=\color{gray},
  keywordstyle=\color{blue}\fontsize{7}{7}\ttfamily,
  basicstyle=\fontsize{7}{7}\ttfamily,
  stringstyle=\color{red},
  frame=single,
  caption={Show the rewritten SQL for the example in Listing~\ref{motivation_example}.},
  label={rewriting_example_listing},
  morekeywords={IFNULL, WITH},
  numbers=left,          
  numberstyle=\tiny\color{gray},  
  stepnumber=1,
  numbersep=5pt]
WITH cte AS
(SELECT 
    IFNULL(MIN(CASE WHEN ds >= '1014' AND ds < '1016' THEN cnt END) / 
             SUM(CASE WHEN ds = '1016' THEN 1 ELSE 0 END), 100) AS dcrs
FROM (
    SELECT ds, COUNT(*) AS cnt
    FROM tb0
    WHERE ds >= '1014' AND ds <= '1016'
    GROUP BY ds
)
SELECT tb0.c0, 
(SELECT tb3.c1 - tb3.c2 FROM tb3 WHERE tb3.ds = tb1.ds),
(SELECT AVG(tb4.c3) FROM tb4 WHERE tb4.ds = tb1.ds AND tb4.c3 > 100)
FROM tb1 
INNER JOIN tb2 ON tb1.ds = tb2.ds 
WHERE tb1.dcrs <= (SELECT dcrs FROM cte)
\end{lstlisting}

\subsection{Equivalence Verifier}

We conduct a systematic review of representative SQL equivalence verification methods and summarize their characteristics in Table~\ref{equivalence_literature}. 

Formal methods such as SPES~\cite{zhou2022spes} offer rigorous correctness guarantees through symbolic execution but are limited in practical applicability due to type constraints and poor performance on real-world benchmarks like TPC-H~\cite{tpch}. Graph-based approaches (e.g., FuncEvalGMN~\cite{FuncEvalGMN}) achieve broader coverage via structural matching but require extensive model training, leading to high deployment costs and limited generalization across diverse query patterns. LLM-based solutions~\cite{zhao2023llmsqlsolver} employ advanced prompting techniques but still suffer from limited accuracy on complex queries and exhibit positive bias in equivalence judgments~\cite{wu2024metarewarding}.

\begin{table}[htbp!]
\centering
\vspace{-2.5mm}
\caption{Comparison of representative SQL equivalence verification methods.}
\vspace{-2.5mm}
\begin{adjustbox}{width=0.95\columnwidth}
{
\begin{tabular}{lcccc}
\toprule
\textbf{Method} & \makecell[c]{\textbf{Technical} \\ \textbf{Basis}} & \makecell[c]{\textbf{Correctness} \\ \textbf{Guarantee}} & \makecell[c]{\textbf{Applicable} \\ \textbf{Scope}} & \makecell[c]{\textbf{Deployment} \\ \textbf{Cost}} \\
\midrule
SPES~\cite{zhou2022spes}                    & Symbolic Execution    & \checkmark    & Very Limited & Low \\
SQLSolver~\cite{ding2023SQLSolver}          & Formal Logic          & \checkmark    & Limited & Low \\
FuncEvalGMN~\cite{FuncEvalGMN}              & Graph Matching        & $\times$      & General & High \\
LLM-SQL-Solver~\cite{zhao2023llmsqlsolver}  & Probabilistic LM      & $\times$      & General & Medium \\
\ours                                       & Probabilistic LM      & $\times$      & \texttt{SELECT}-based DML  & Medium \\
\bottomrule
\end{tabular}
}
\end{adjustbox}
\vspace{-2.5mm}
\label{equivalence_literature}
\end{table}

To address these limitations, we propose a structured framework for SQL semantic equivalence verification, specifically tailored for \texttt{SELECT}-based DML queries in OLAP scenarios. Our approach decomposes the verification process into two stages: (1) semantic intent extraction, which captures the meaning of each query using controlled LLM-mediated interpretation, and (2) hierarchical consistency checking, which performs field-level alignment and derivation trace analysis to assess equivalence.

In the first stage, each SQL query is translated into a structured natural language representation that captures the provenance of every field in the main \texttt{SELECT} clause. This includes source tables, transformation logic (e.g., aggregations or arithmetic expressions), and filtering or joining conditions. To enhance interpretability, we adopt a stepwise parsing strategy that starts with the innermost subqueries and progressively builds up the outer query semantics.

A lightweight pre-filtering module eliminates clearly inconsistent query pairs based on schema-level heuristics, such as mismatches in field count or source tables. For remaining pairs, an LLM agent performs detailed consistency analysis by aligning corresponding \texttt{SELECT} fields and reasoning about semantic equivalence.

Our prompt design enforces bidirectional field mapping, supports counterexample generation for non-equivalent pairs, and incorporates calibrated confidence scoring—ensuring both the interpretability and reliability of the verification process.

\subsection{Query Modifier}

\begin{figure}[htbp!]
    \centering
    \includegraphics[width=0.95\linewidth]{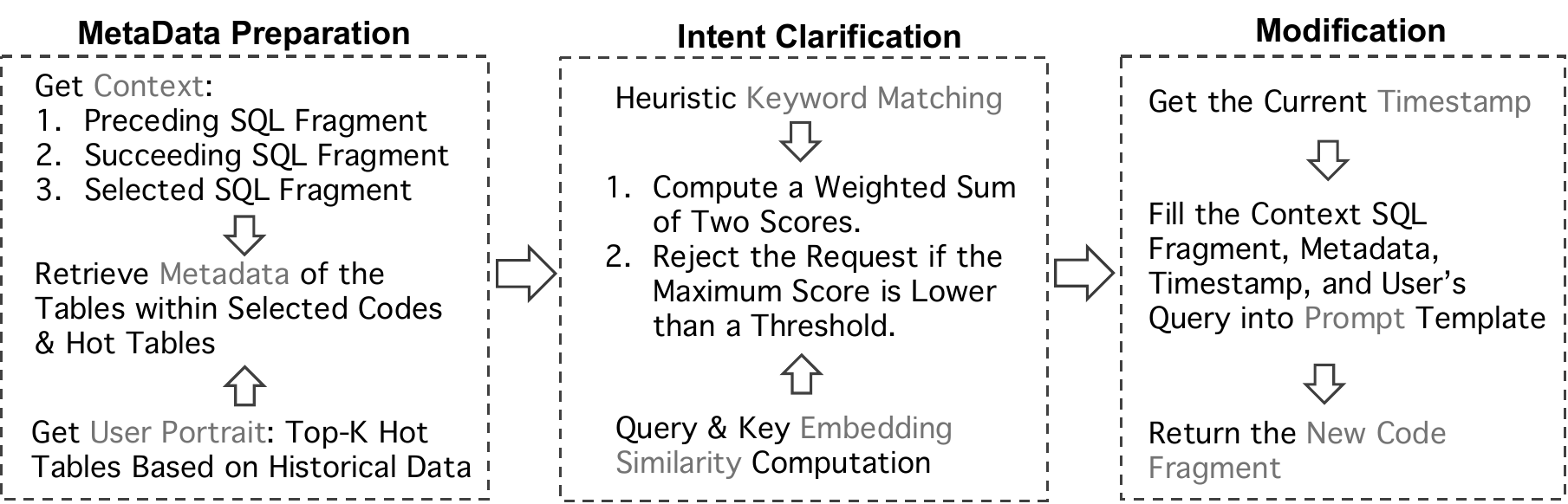}
    \vspace{-2.5mm}
    \caption{Workflow of the query modification process.}
    \vspace{-2.5mm}
  \label{modification}
\end{figure}

We classify modification requests into four general categories:
(1) \textit{Realizing a specified semantics}: Adjust the SQL logic to align with the user’s intended meaning.
(2) \textit{Explaining the SQL}: Preserve the original logic while adding comments or annotations for clarity.
(3) \textit{Adopting a specified syntax}: Maintain semantic equivalence while adapting the query to match the user's stylistic or structural preferences.
(4) \textit{Other SQL-related tasks}: Capture queries that do not clearly fall into the above three categories, such as stylish polishing.
To fulfill each request, the tool follows a three-step pipeline (see Figure~\ref{modification}):
(1) metadata preparation,
(2) user intent clarification, and
(3) query modification. 
Note that the definitions of categories are rather flexible and can be customized.

\subsubsection{Metadata Preparation} 
During the metadata preparation stage, we extract the target SQL snippet along with its surrounding context to provide a comprehensive view of the query environment.
We gather metadata from two primary sources. The first includes tables and columns referenced in the SQL snippet, along with their names and descriptions. The second source is derived from the user's historical query logs, where we identify the top-$k$ most frequently accessed tables. For each of these tables, we extract relevant metadata—such as schema information and usage patterns—to help the LLM better understand the context and semantics of the query.
Additionally, we append a current timestamp to the metadata to provide temporal grounding, which is particularly useful when handling evolving schema or time-sensitive queries.

\subsubsection{User Intent Clarification} 
The user intent clarification step maps the natural language request to one of the four predefined categories described earlier. This classification combines two complementary strategies: heuristic keyword matching and semantic similarity scoring.
In the heuristic keyword matching strategy, we identify a set of domain-specific keywords and phrases that are commonly associated with each modification type. 
For each category \( C_j \), we define a corresponding keyword set \( \mathcal{KW}_j = \{k_{j1}, k_{12}, \ldots, k_{jn_j}\} \). Given a user request \( \mathcal{Q} \), we compute a weighted matching score \( S_j^{kw} \) for each category as follows:
\begin{equation}
S_j^{kw} = \frac{1}{N_j} \sum_{k_{ji} \in \mathcal{KW}_j} \mathrm{match}(\mathcal{Q}, k_{ji}) \times w_{ji}
\end{equation}
where
$N_j$ is the total number of candidate keywords in $\mathcal{KW}_j$; \(\mathrm{match}(\mathcal{Q}, k_{ji})\) is a binary function returning 1 if the keyword \( k_{ji} \) appears in the request \( \mathcal{Q} \), and 0 otherwise; \( w_{ji} \) denotes the weight assigned to keyword \( k_{ji} \), reflecting its relative importance within category \( C_j \).

To complement the keyword-based method, we also employ semantic embeddings to capture more nuanced intent signals. We have explored two distinct embedding pathways:
(1) Sentence-Transformer with Masking: We pre-process the query by replacing specific metadata (e.g., table/column names) and constant values with special tokens \texttt{[MASK]}, then encode the masked text into a vector $\mathbf{e}_{\mathcal{Q}}$ using a Sentence-Transformer model (e.g., RoBERTa~\cite{liu2019roberta}). 
(2) Instruction‑aware Qwen3‑Embedding: We utilize the Qwen3 embedding model~\cite{zhang2025qwen3} to encode the original query $\mathcal{Q}$, guided by an instruction prompt that directs the model to focus on the user's action intent rather than details such as schema identifiers. 
During our development and testing, the Instruction-Aware Qwen3 Embedding consistently outperforms the masking-based Sentence-Transformer approach in both classification accuracy and robustness to domain variations. This is attributed to its ability to better align with the LLM’s internal reasoning process and its use of instruction-tuned representations that emphasize action-oriented semantics. What's more, real-world user queries often suffer from ambiguity, incomplete descriptions, or informal phrasing. In such cases, rule-based detection and masking strategies tend to fail.

Once the embedding method has been selected, we construct a representative embedding vector $\mathbf{e}_{\mathcal{C}_j}$ for each category \( C_j \). These are derived by collecting historical user queries, classifying their intents using an LLM, and computing the centroid embedding for each category. 

We employ cosine similarity between the query embedding $\mathbf{e}_{\mathcal{Q}}$ and the category centroid embeddings $\mathbf{e}_{\mathcal{C}_j}$ as a measure of semantic proximity. This semantic score is combined with the heuristic keyword matching score to form the final classification decision. 
Formally, we define the final classification score \( F_j \) for each category $\mathcal{C}_j$ as a weighted combination:
\begin{equation}
F_j = \alpha \cdot S_j^{kw} + \beta \cdot \text{similarity}(\mathbf{e}_Q, \mathbf{e}_{\mathcal{C}_j})
\end{equation}
where:
\( \alpha \) and \( \beta \) are weighting parameters used to balance the heuristic keyword matching score and the semantic similarity score. 
To ensure robust and reliable intent clarification, we introduce a confidence threshold \( \theta \). If the maximum classification score \( \max(F_j) \) across all categories falls below this threshold, the system treats the request as unsupported and rejects the modification task. This mechanism helps filter out ambiguous or outlier queries that do not align well with any known modification type, thereby maintaining the integrity and reliability of the classification pipeline.

\subsubsection{Modification} 
Once the necessary data has been collected and the user's intent has been classified, we construct a structured prompt tailored to the identified modification type. The prompt integrates the following components: the original SQL fragment, its surrounding context, relevant metadata (e.g., schema information and top accessed tables), the current timestamp for temporal grounding, and the natural language instruction from the user. This contextualized prompt is then fed into the LLM, which reasons over the input and generates a modified SQL fragment that accurately fulfills the user's intent while preserving correctness and consistency.

\subsection{Syntax Error Corrector}

\begin{figure}[htbp!]
    \centering
    \includegraphics[width=0.95\linewidth]{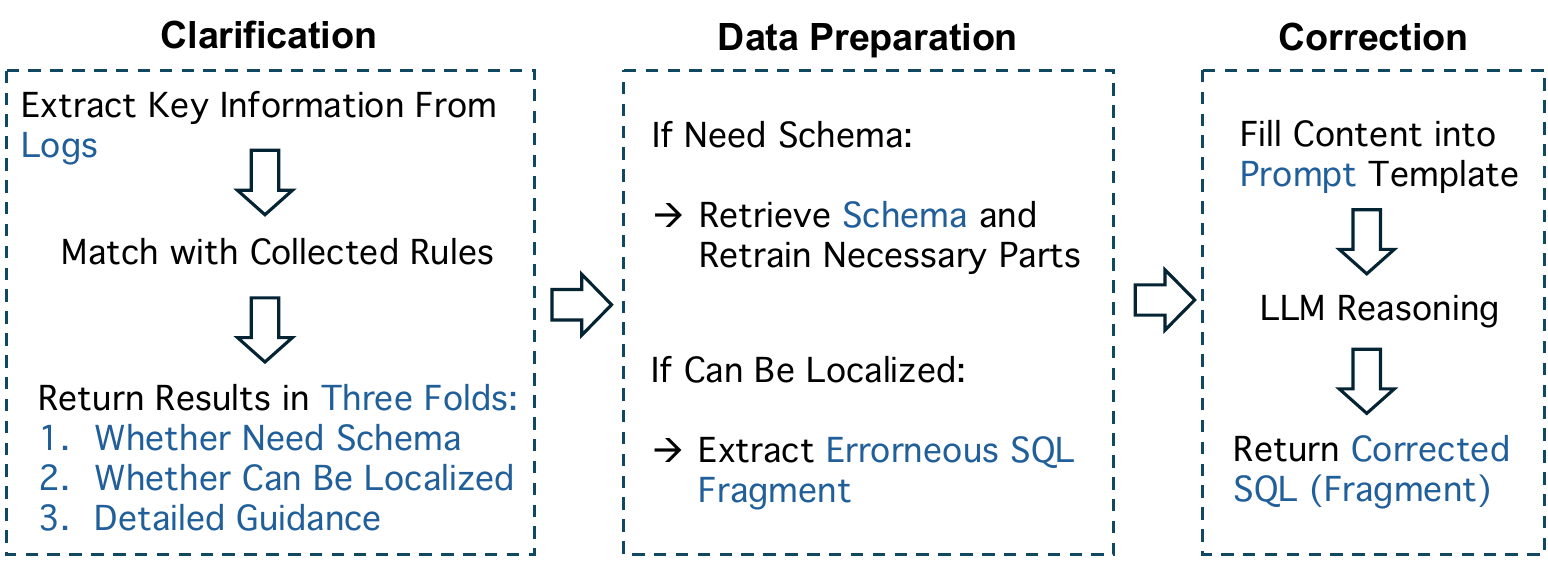}
    \vspace{-2.5mm}
    \caption{Workflow of the syntax error correction process.}
    \vspace{-2.5mm}
  \label{correction}
\end{figure}

The overall syntax correction workflow is illustrated in Figure~\ref{correction}, comprising three key stages: clarification, data preparation, and correction.

\subsubsection{Clarification} 
The clarification stage begins with the input of complete error logs generated by the DBMS when executing a syntactically incorrect SQL query. From these logs, we extract structured information—such as the exception type, error location, and descriptive message—using regular expression matching. This extracted information is then used to perform embedding-based retrieval against the knowledge base of known error patterns and corresponding correction strategies. Each retrieved strategy includes three components: (1) Whether schema information is required to correction; (2) Whether the error can be resolved through localized correction; (3) Detailed guidance on how to address the error. 

The first component addresses the varying degrees of schema dependency across different error types. For example, syntax errors involving column names often require access to schema metadata, whereas errors related to SQL keywords or built-in functions may not. In OLAP scenario, database schema can be large and complex~\cite{LiLCLT24/nl2sql360}, with tables containing dozens of columns with lengthy names. 

The second component determines whether an error is localized or global in nature. Some syntax issues are confined to small fragments of the query and can be corrected without considering the entire structure. For instance, a missing comma between column names is a localized error that can be fixed by inspecting only the immediate context. Based on this, we classify errors in local or global categories and adopt corresponding correction strategies.

The third component provides explicit correction hints, which serve as actionable guidance for the LLM agent. These hints direct the model's reasoning process toward effective solutions. 
For example, in the case of a \texttt{Column count mismatch} error, a useful hint would be: 
``This error occurs when \texttt{SELECT} clauses connected by \texttt{UNION} or \texttt{UNION ALL} contain a different number of fields.'' 
Such contextualized instructions help the LLM better understand the nature of the error and apply domain-specific correction strategies accordingly. 

Importantly, the decisions regarding localized correction scope and selective schema inclusion aim to keep the prompt concise. This reduces both inference latency and computational cost while avoiding the inclusion of irrelevant information that could interfere with the model's reasoning capability~\cite{ICPCPromptCompression}.

\subsubsection{Data Preparation}
Guided by the outputs from the clarification stage, the data preparation step determines what information should be included in the final prompt. It selectively extracts relevant schema components, or isolates specific query fragments for localized correction, depending on the retrieved strategy. If the clarification stage fails to find a confident match in the knowledge base, a conservative fallback strategy is applied: the full schema is retained, and a global correction approach is used.

\subsubsection{Correction}
In the final correction stage, the erroneous SQL (or fragment), along with its associated context—including selectively extracted schema information and tailored guidance—is assembled into a structured prompt. The LLM agent then processes this input and generates a corrected version of the SQL query.

%% file: experiment.tex
\section{Experiments}

\subsection{Datasets and Metrics}

\noindent\textbf{Datasets}. To comprehensively evaluate the performance of \ours, we select two representative benchmarks: BIRD-CRITIC~\cite{li2025swe} and BIRD~\cite{li2024bird}. Additionally, we have constructed a new dataset named Payment-SQL, which comprises analytical SQL queries derived from real industrial scenarios, specifically designed to evaluate performance in handling complex and diverse queries.

\textbf{BIRD-CRITIC} is an innovative SQL benchmark crafted to evaluate the critical capabilities of LLMs in diagnosing and resolving user issues within real-world database environments. The benchmark categorizes issues into four domains: \textit{Query}, \textit{Management}, \textit{Personalization}, and \textit{Efficiency}. These categories align with the core functionalities of \ours. For our experiments, we utilize a light version, \texttt{bird-critic-1.0-flash-exp}, which consists of 200 user issues on PostgreSQL.

\begin{table}[htbp!]
  \centering
  \vspace{-2.5mm}
  \caption{Evaluation on BIRD’s dev set.}
  \vspace{-2.5mm}
  \begin{tabular}{l p{1.8cm} p{1.8cm}}
    \toprule
    \multirow{2}{*}{\makecell{\textbf{Methods}}} & \multicolumn{2}{c}{\textbf{Dev set}} \\
    & \textbf{EX(\%)} & \textbf{VES(\%)} \\
    \midrule
    \rowcolor{my_lightgray} \multicolumn{3}{c}{\textbf{Prompt-based base models}} \\
    \midrule
    Codex~\cite{li2024bird} & 34.35 & 43.41 \\
    ChatGPT~\cite{li2024bird} & 37.22 & 43.81 \\
    GPT-4~\cite{li2024bird} & 46.35 & 49.77 \\
    DIN-SQL + GPT-4~\cite{dinsql} & 50.72 & 58.79 \\
    DAIL-SQL + GPT-4~\cite{dailsql} & 54.76 & 56.08 \\
    \midrule
    \rowcolor{my_lightgray} \multicolumn{3}{c}{\textbf{Fine-tuning-based base models}} \\
    \midrule
    T5-3B~\cite{li2024bird} & 23.34 & 25.57 \\
    CodeS-7B~\cite{codes} & 57.17 & 58.80 \\
    CodeS-15B~\cite{codes} & 58.48 & 59.87 \\
    XiYan-32B~\cite{xiyansql} & 67.01 & 67.79 \\
    \midrule
    \rowcolor{my_yellow} \multicolumn{3}{c}{\textbf{With post-processing tools}} \\
    \midrule
    CodeS-7B + SQLFixAgent~\cite{cen2024sqlfixagent} & 60.17 (\textuparrow3.00) & 63.15 (\textuparrow4.35) \\
    \midrule
    CodeS-7B + \ours  & 64.02 (\textbf{\textuparrow6.85}) & 64.72 (\textuparrow5.92) \\
    CodeS-15B + \ours & 65.32 (\textuparrow6.84) & 67.87 (\textbf{\textuparrow8.00}) \\
    XiYan-32B + \ours & \textbf{68.97} (\text{\textuparrow1.96}) & \textbf{70.89} (\textuparrow3.10) \\
    \bottomrule
  \end{tabular}
  \vspace{-2.5mm}
  \label{tab:bird_performance}
\end{table}

\textbf{BIRD} serves as a challenging large-scale database text-to-SQL evaluation benchmark, designed to bridge the gap between academic research and practical applications. It encompasses 95 extensive databases and high-quality text-SQL pairs, with data storage reaching up to 33.4GB, spanning 37 professional fields. The validation set includes 1,534 test entries, offering a comprehensive evaluation of text-to-SQL translation capabilities. Notably, in utilizing this dataset, we employ \ours as a post-processing tool for NL2SQL models, aimed at further enhancing the quality of generated SQL queries.

\textbf{Payment-SQL dataset} originates from real-world industrial OLAP scenarios and is curated by human experts based on execution logs. It contains 50 SQL queries, each involving an average of 2 tables and 11 columns, drawn from a schema of 74 tables with thousands of fields. Designed specifically for evaluating SQL rewriting systems, Payment-SQL measures effectiveness through execution time comparisons before and after rewriting in the same environment—directly reflecting real-world performance gains. A key feature of Payment-SQL is its complexity: the average query length is 421 tokens , far exceeding that of BIRD’s challenging category (107 tokens). According to Spider 2.0~\cite{spider2.0}, where queries over 160 tokens are considered difficult, even the shortest query in Payment-SQL (173 tokens) qualifies as hard, with the longest reaching 1169 tokens. This makes Payment-SQL a rigorous and realistic benchmark for evaluating the robustness and scalability of SQL rewriting techniques in industrial applications. The dataset is available at \url{https://anonymous.4open.science/r/SQLGovernor-33DF}.

\begin{table*}[htbp]
\centering
\vspace{-2.5mm}
\caption{Evaluation of \ours on BIRD-CRITIC-Flash. Metric: SR (\%).}
\vspace{-2.5mm}
\resizebox{0.85\linewidth}{!}{
\begin{tabular}{cccc|cc|cc|cc|cc}
\toprule
\multicolumn{2}{c}{\multirow{2}{*}{\textbf{Method}}} & \multicolumn{10}{c}{\textbf{Category}} \\ 
\cline{3-12}
 & & \multicolumn{2}{c|}{Query} & \multicolumn{2}{c|}{Management} & \multicolumn{2}{c|}{Personalization} & \multicolumn{2}{c|}{Efficiency} & \multicolumn{2}{c}{Total} \\ 
\hline
\multirow{2}{*}{\makecell[c]{\textbf{Qwen3-32B}}} & / & 18.8 & / & 34.0 & / & 28.1 & / & 22.7 & / & 26.0 & /\\ 
\cline{3-12}
 & + \ours & 21.9 & \textuparrow 3.1 & \textbf{54.0} & \textbf{\textuparrow 20.0} & 35.9 & \textuparrow 7.8 & 36.4 & \textbf{\textuparrow 13.7} & 36.0 & \textbf{\textuparrow 10.0} \\ 
\hline 
\multirow{2}{*}{\makecell[c]{\textbf{Qwen2.5} \\ \textbf{-72B-Instruct}}} & / & 20.3 & / & 46.0 & / & 32.8 & / & 31.8 & / & 32.0 & / \\ 
\cline{3-12}
 & + \ours & \textbf{26.6} & \textbf{\textuparrow 6.3} & 52.0 & \textuparrow 6.0 & \textbf{43.8} & \textbf{\textuparrow 11.0} & \textbf{45.5} & \textuparrow 13.6 & \textbf{40.5} & \textuparrow 8.5 \\ 
\bottomrule
\end{tabular}
}
\vspace{-2.5mm}
\label{bird-critic}
\end{table*}

\textbf{Evaluation Metrics}.
On the BIRD-CRITIC-FLASH dataset, we follow the official guidelines and use the success rate (SR) as the metric, as it effectively evaluates multiple aspects of performance due to the well-designed test cases. For the BIRD dataset, we employ both Execution Accuracy (EX) and Valid Efficiency Score (VES) metrics to comprehensively evaluate performance. In the case of the Payment-SQL dataset, rewriting effectiveness is assessed using Execution Time Saved (ETS) and Execution Time Optimization Gain (ETOG), calculated as follows:
\begin{equation}
\text{ETS} = \text{ET}_{\text{pre}} - \text{ET}_{\text{post}},\ 
\text{ETOG} = \frac{\text{ETS}}{\text{ET}_{\text{pre}}  }\times 100\%,
\end{equation}
where $ET_{\text{pre}}$ represents the execution time before rewriting and $ET_{\text{post}}$ represents the execution time after rewriting. It is worth noting that when using ETS and ETOG to evaluate SQL rewriting tasks, we typically execute both the pre-optimized and post-optimized SQL queries in the same system while excluding interference factors such as execution caching to ensure the objectivity and reliability of the test results.

\subsection{Main Results}
\label{main_results}

\subsubsection{Results on BIRD}

As \ours is used as post-processing tool for the NL2SQL task, we select three strong base models and one representative baseline. The base models are CodeS-7B, CodeS-15B~\cite{codes} and XiYanSQL~\cite{xiyansql} and the baseline is \textit{SQLFixAgent}~\cite{cen2024sqlfixagent}.

\begin{figure}[htbp!]
    \centering
    \begin{subfigure}[b]{0.47\linewidth}
        \centering
        \includegraphics[width=\linewidth]{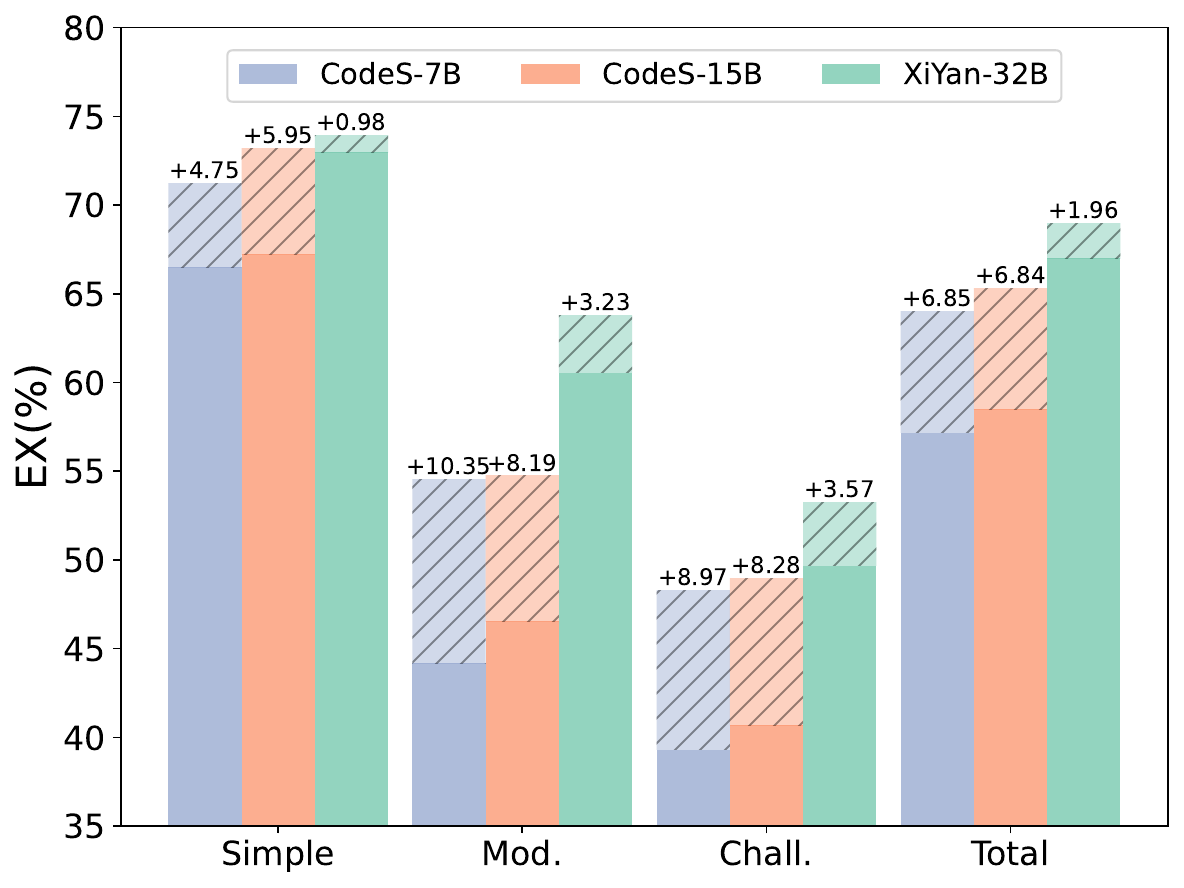}
        \caption{}
        \label{fig:diff_category}
    \end{subfigure}
    \hfill
    \begin{subfigure}[b]{0.47\linewidth}
        \centering
        \includegraphics[width=\linewidth]{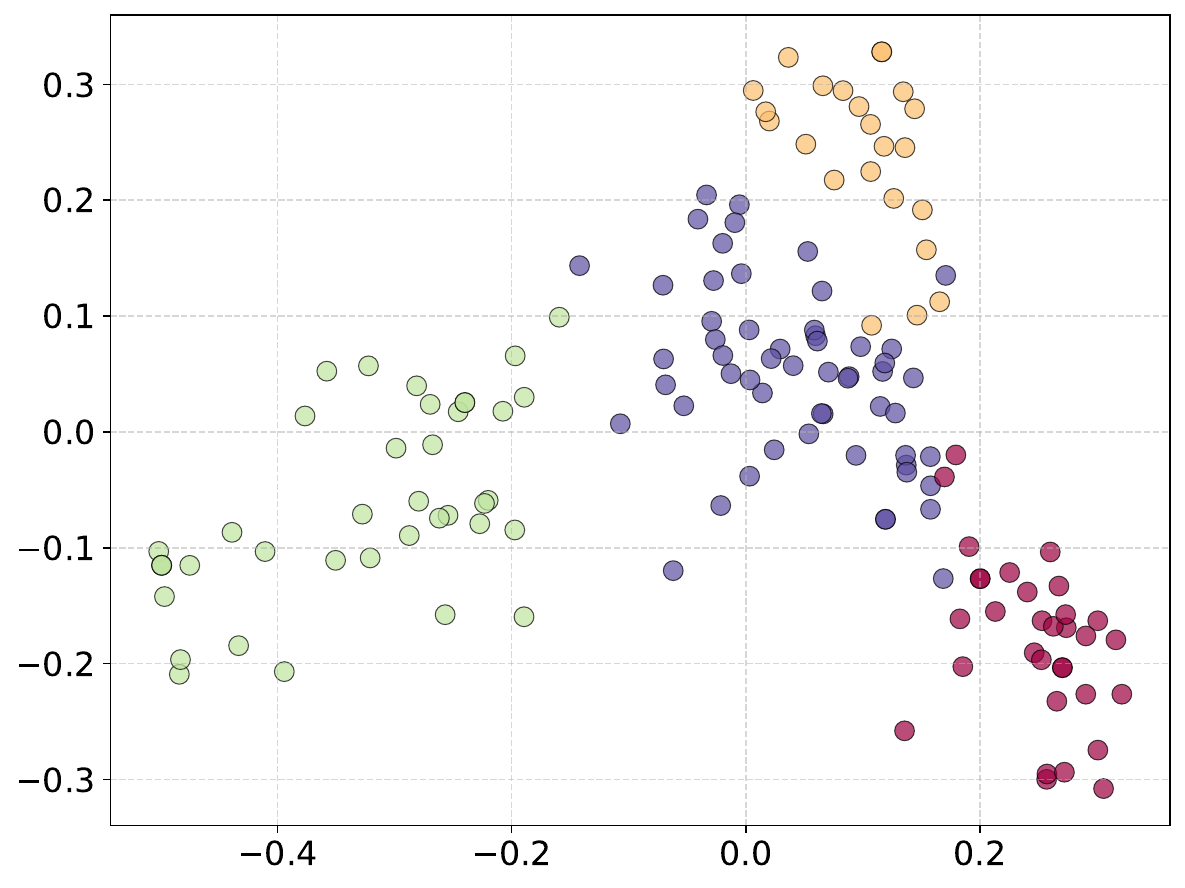}
        \caption{}
        \label{intent_cluster}
    \end{subfigure}
    \vspace{-2.5mm}
    \caption{(a) Performance on various difficulty data categories. (b) Visualization results of user query clustering.}
    \vspace{-4.5mm}
    \label{category_diff_user_intent}
\end{figure}

Table~\ref{tab:bird_performance} presents a detailed comparison of \ours's performance against existing methods on the BIRD dataset. 
When integrated with CodeS-15B, \ours achieves EX and VES scores of 65.32\% and 67.87\%, respectively, representing improvements of 6.84\% in EX and 8.00\% in VES over the baseline. These gains outpace those of SQLFixAgent, which improves the baseline by only 3\% in EX and 4.35\% in VES, highlighting \ours's superior effectiveness.
When paired with XiYan-32B, \ours further enhances the already high baseline scores of 68.97\% (EX) and 70.89\% (VES), achieving marginal but meaningful improvements of 1.96\% in EX and 3.10\% in VES.

Furthermore, we carefully analyze the detailed performance of \ours across various base models. Figure ~\ref{fig:diff_category} presents the results of this method on the three difficulty levels (\texttt{Simple}, \texttt{Mod.}, and \texttt{Chall.}) of the BIRD dev set. The solid bars represent the results of the base model, while the dashed bars above indicate the gains achieved by \ours. It is evident that \ours achieves significantly higher gains on the \texttt{Mod.} and \texttt{Chall.} difficulty levels compared to Simple. Notably, on the CodeS-15B and XiYan-32B models, the metric gains for \texttt{Chall.} even surpass those for Mod., making it the highest-performing category among the three difficulty levels, with respective gains of +8.28\% and +3.57\%. This clearly demonstrates the advantage of \ours in handling long and complex SQL queries.

In summary, \ours exhibits strong performance improvements across all tested baseline models, outperforming alternative methods such as SQLFixAgent and demonstrating promising value even when applied to state-of-the-art models like XiYan-32B.

\subsubsection{Results on BIRD-CRITIC-Flash}

To evaluate the effectiveness of \ours in addressing SQL issues arising from user-provided natural language queries, we conduct experiments on the BIRD-CRITIC-Flash benchmark. Our approach dynamically routes each SQL issue to the most suitable tool: the Query Rewriter for efficiency-related problems, the Syntax Error Corrector for execution errors, and the Query Modifier for other semantic or stylistic adjustments. 

We evaluate two widely used LLMs—Qwen3-32B and Qwen2.5-72B-Instruct—in both the original configurations provided by the benchmark team and with our toolkit integrated. As shown in Table~\ref{bird-critic}, \ours consistently improves performance across all categories and both base models.

For Qwen3-32B, integrating \ours leads to substantial gains, particularly in the Management category (+20.0\%), where the success rate increases from 34.0\% to 54.0\%. Significant improvements are also observed in Efficiency (+13.7\%) and Personalization (+7.8\%), indicating that both query rewriting and semantic alignment benefit greatly from our toolkit. Overall, the total score rises from 26.0\% to 36.0\%. When applied to Qwen2.5-72B-Instruct, \ours still delivers consistent gains. The largest improvement is seen in Personalization (+11.0\%).

In addition to success rate, we measure the average end-to-end inference time per query on both models. For Qwen3-32B, the runtime increases from 8.5s (base) to 18.4s with our toolkit; similarly, for Qwen2.5-72B-Instruct, it increases from 9.8s (base) to 21.3s. While this represents a non-trivial overhead, it is largely due to the multi-stage processing pipeline—including intricate problem analysis and solving process powered by LLM-that is essential for achieving high-quality corrections in complex OLAP queries.

\subsection{Ablation Study}

\subsubsection{User Intent Clarification}

To evaluate the effectiveness of user intent clarification in the SQL Modifier, we sampled 150 real-world query tasks from our production environment and manually annotated them with intent categories. We then tested the performance of the Instruction-Aware Qwen3 Embedding in classifying these intents. Specifically, we used an embedding model with 8B parameters and a vector dimension of 1024. As a baseline, we also evaluated Qwen3-32B, where the LLM directly performs classification without prior embedding-based filtering. Both models were deployed under identical execution environments to ensure fair comparison. 

The results are as follows: the Instruction-Aware Qwen3 Embedding achieves an accuracy of 78.9\%, with an average inference latency of 0.173 seconds. In contrast, Qwen3-32B achieves higher accuracy at 84.3\%, but incurs a significantly higher average latency of 0.354 seconds. These findings suggest that while the LLM-based classifier offers relatively higher accuracy (5.4\%), the embedding-based approach provides a favorable trade-off between speed and performance—making it particularly suitable for high-throughput or latency-sensitive applications.

To provide a more intuitive understanding of the embedding quality, we applied PCA to reduce the embedding vectors to two dimensions and visualized them using scatter plots, as shown in Figure~\ref{intent_cluster}.

\subsubsection{Error Correction}

To evaluate the error correction capabilities of \ours, we collect a set of syntactically and semantically incorrect SQL queries generated by two strong LLM-based NL2SQL systems—CodeS-7B and CodeS-15B—on the BIRD dataset. Queries that failed to execute due to syntax errors are fed into the Syntax Error Corrector, while those exhibiting semantic misalignment are routed to the Query Modifier for refinement.

Table~\ref{tab:dataset_error_case} presents the results of the error correction capabilities in \ours. The findings indicate that the module demonstrates strong error correction performance on the BIRD datasets, as evidenced by the predictive results from both baseline models.
For the CodeS-7B model, we analyzed 691 erroneous cases, yielding an overall EX rate of 25.8\%. Performance across difficulty levels shows EX rates of 26.4\% for simple cases, 26.2\% for moderate cases, and 22.0\% for challenging cases. In contrast, the CodeS-15B model, evaluated on 667 erroneous cases, achieved an overall EX rate of 25.2\%, with rates of 26.9\% for simple cases, 23.1\% for moderate cases, and 25.0\% for challenging cases.

\begin{table}[htbp!]
    \centering
    \caption{Error correction performance on the BIRD's dev set including syntactic and semantic levels.}
    \vspace{-2.5mm}
    \begin{tabular}{lcccc}
    \toprule
        \textbf{Error Data Statistics} & \textbf{Total} & \textbf{Simple} & \textbf{Mod.} & \textbf{Chall.} \\
    \midrule
   \textbf{\#CodeS-7B Error Case} & 691 & 333 & 267 & 91 \\
    \textbf{EX(\%)} & 25.8 & 26.4 & 26.2 & 22.0 \\
    \midrule
    \textbf{\#CodeS-15B Error Case} & 667 & 324 & 255 & 88 \\
    \textbf{EX(\%)} & 25.2 & 26.9 & 23.1 & 25.0 \\
    \bottomrule
    \end{tabular}
    \vspace{-2.5mm}
    \label{tab:dataset_error_case}
\end{table}

\subsubsection{Rewriting}

To evaluate the performance of SQL rewriting tool, we used Payment-SQL, a test set that closely aligns with OLAP scenario, and employed ETS and ETOG as evaluation metrics. 
We selected four representative models as baselines: Qwen2.5-72B-Instruct~\cite{qwen2.5}, Qwen3-32B~\cite{zhang2025qwen3}, LLM-R$^2$~\cite{li2024llm-r2}, and GenRewrite~\cite{liu2024genrewrite}.
Specifically, Qwen2.5-72B-Instruct and Qwen3-32B are general-purpose LLMs that are instructed to rewrite the input SQL query in a single inference step. LLM-R$^2$ employs LLMs to select appropriate rewriting rules and trains a separate demonstration recommendation model to guide the rewriting process. GenRewrite represents the first non-rule-based, end-to-end query rewriting approach that fully leverages the generative capabilities of LLMs.

It is worth noting that we deliberately exclude conventional rule-based pattern-driven approaches. These methods treat query rewriting as a direct transformation between execution plans, so they are orthogonal to our design.
To ensure fair comparisons, we maintained identical execution environments for the SQL queries before and after rewriting during testing.

\begin{table}[hbtp!]
    \centering
    \caption{Execution efficiency of SQLs in Payment-SQL after rewriting and time-cost for rewriting.}
    \vspace{-2.5mm}
    \begin{tabular}{ccccc}
    \toprule
       \textbf{Methods} & \textbf{ETOG(\%)} & \textbf{ETS(s)} & \textbf{Cost(s)} & \textbf{$\Delta$(s)} \\
       \hline
       Qwen3 & 11.06  & 20.19 & 15.24 & \textuparrow 4.95 \\
        Qwen2.5 & 14.56 & 26.59 & 18.41 & \textuparrow 8.18 \\
        LLM-R$^2$ & 29.87 & 54.53 & 46.85 & \textuparrow 7.68 \\
        GenRewrite & 31.25 & 57.07 & 38.36 & \textuparrow 18.71 \\
        \ours & \textbf{45.92} & \textbf{83.86} & 30.73 & \textbf{\textuparrow 53.13} \\
    \bottomrule
    \end{tabular}
    \vspace{-2.5mm}
    \label{tab:sql_opti_wxpay}
\end{table}

The results are presented in Table~\ref{tab:sql_opti_wxpay}. From the table, we observe that \ours exhibits a significant performance advantage when applied to industrial-level OLAP workloads. On average, across the entire test set, \ours achieves a 45.92\% reduction in execution time and an 83.86-second reduction in absolute execution time. We also report the rewriting cost, i.e., the time required to perform the rewriting itself, and the net benefit ($\Delta$), defined as the difference between ETS and Cost. Notably, while \ours incurs a relatively moderate rewriting overhead (30.73s), it delivers the largest net benefit (+53.13s), demonstrating its practical viability in real-world applications where query latency is critical.

Listing~\ref{rewriting_example} presents an example of an rewritten SQL query from Payment-SQL. During the evaluation stage, the LLM provided the following rewriting suggestions: 
(1) Use the \texttt{WITH} clause to explicitly define the result of \texttt{UNION ALL} as a temporary table, making it easier for the rewriter to understand and optimize the query. 
(2) In the \texttt{UNION ALL} step, explicitly select only the columns that are actually needed, avoiding the retrieval and processing of unnecessary data.
\begin{lstlisting}[
  language=SQL, 
  commentstyle=\color{gray},
  keywordstyle=\color{blue}\fontsize{7}{7}\ttfamily,
  basicstyle=\fontsize{7}{7}\ttfamily,
  stringstyle=\color{red},
  frame=single,
  caption={Example of rewriting result from Payment-SQL.},
  label={rewriting_example},
  morekeywords={OVER, PARTITION, WITH}]
-- Original SQL
SELECT AVG(duration)
FROM (
    SELECT *, row_number() OVER (PARTITION by instanceid ORDER BY modifytime DESC) AS id
    FROM (
        SELECT *
        FROM table0 
        WHERE ds > '0201'
        UNION ALL
        SELECT *
        FROM table1 
        WHERE ds > '0201'
    ) a
) b
WHERE id = 1 AND taskid IN (1, 12, 123) AND scriptid = 666 
-- Rewritten SQL
WITH combined_data AS (
    SELECT taskid, instanceid, scriptid, modifytime
    FROM table0
    WHERE ds > '0201'
    UNION ALL
    SELECT taskid, instanceid, scriptid, modifytime
    FROM table1
    WHERE ds > '0201'
)
SELECT AVG(duration)
FROM
(
    SELECT *, ROW_NUMBER() OVER (PARTITION BY instanceid ORDER BY modifytime DESC) AS id
    FROM combined_data
) b
WHERE id = 1 AND taskid IN (1, 12, 123) AND scriptid = 666
\end{lstlisting}

\subsubsection{Equivalence Verification}

We use the predictive results from the CodeS-7B model alongside golden SQL queries to establish positive and negative pairs. Correctly predicted SQL queries are classified as equivalent with the golden SQL (labeled as true), while incorrectly predicted queries are deemed nonequivalent (labeled as false). The results are presented in Table~\ref{tab:dataset_verif}. 
We report two metrics-accuracy and F1 score-and present the results in Table~\ref{tab:dataset_verif}, the overall accuracy for verification is 78.9\% and F1 score is 79.3\%, indicating effective performance of the Verifier. It is noteworthy that the scores for challenging queries are lower than those for simpler queries, which is expected given the increased complexity of the SQL statements.

\begin{table}[htbp!]
    \centering
    \caption{Equivalence verification performance on the predictive results of CodeS-7B.}
    \vspace{-2.5mm}
    \begin{tabular}{lcccc}
    \toprule
        \textbf{Data Category} & \textbf{Total} & \textbf{Simple} & \textbf{Mod.} & \textbf{Chall.} \\
    \midrule
    \textbf{Verif. Accuracy(\%)} & 78.9 & 81.2 & 76.9 & 71.0 \\
    \textbf{Verif. F1(\%)} & 79.3 & 84.3 & 69.9 & 57.1 \\
    \bottomrule
    \end{tabular}
    \vspace{-2.5mm}
    \label{tab:dataset_verif}
\end{table}

\subsection{Detailed Analysis}

\subsubsection{Capability in processing long and complex SQL}

We analyze the capability of \ours in handling long SQL queries from two common scenarios: error correction and rewriting. Figure~\ref{fig:diff_category} presents the detailed performance of \ours when using CodeS-7B, CodeS-15B, and XiYan-32B as base models. The shaded bars illustrate the performance improvement achieved by \ours over the base models. \ours consistently outperforms the base models across all categories, with particularly notable gains in the Challenge SQL section of the BIRD dataset.

Furthermore, Table~\ref{tab:sql_opti_wxpay} illustrates the rewriting performance of \ours on the industrial-level dataset Payment-SQL. Compared to general-purpose LLMs, \ours exhibits a clear advantage in SQL rewriting tasks. Notably, the average token length of Payment-SQL reaches 421, far exceeding the complexity of SQL queries in the BIRD dataset. Additionally, all SQL queries in Payment-SQL meet the Hard SQL standard defined by Spider 2.0~\cite{spider2.0} (token length > 160). These results strongly demonstrate the superior rewriting capability of \ours in handling long and complex SQL queries.

\subsubsection{Validation of productivity improvement}

To validate the effectiveness of \ours in addressing productivity bottlenecks caused by fragmented SQL tool-chains, we conducted a controlled A/B testing with 60 practitioners from in-production data platform. Participants were stratified by SQL expertise (30 experts and 30 non-experts) and uniformly assigned to two groups-Group A: Utilizing the integrated \ours framework; Group B: Operating equivalent discrete modules through manual orchestration. Each subject executed 50 standardized SQL governance tasks spanning evaluation, correction, rewriting, verification. We systematically measured-task completion time and tool-switching frequency. Results demonstrated statistically significant advantages for the integrated framework. Group A achieved 33\% faster task completion, with non-experts exhibiting greater efficiency gains (41\% improvement) compared to 25\% for experts. This disparity correlates with Group B's tool-switching patterns, where practitioners incurred 18\% temporal overhead reconstructing workflow contexts between discrete modules. The empirical evidence quantitatively confirms that \ours's unified pipeline effectively mitigates fragmentation-induced productivity loss, particularly benefiting non-specialist users.

\subsubsection{``Evolving with every step''}

To validate the effectiveness of our expert-guided hybrid self-learning mechanism in continuously enhancing the performance of \ours across various SQL-related tasks, we collect and retain results at different stages for an end-to-end task and an individual tasks. This approach allows us to assess how \ours improves its capabilities through self-learning. 

\begin{figure}[htbp!]
    \centering
    \begin{subfigure}[b]{0.47\linewidth}
        \centering
        \includegraphics[width=\linewidth]{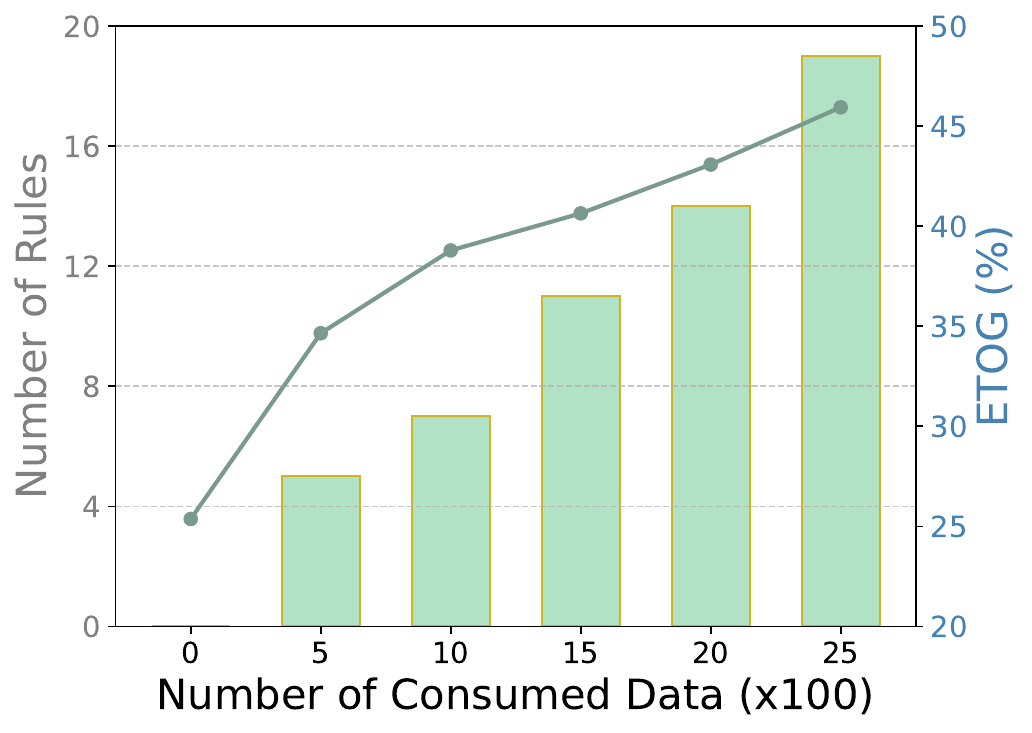}
        \caption{Rewriting}
        \label{payment_sl}
    \end{subfigure}
    \hfill
    \begin{subfigure}[b]{0.47\linewidth}
        \centering
        \includegraphics[width=\linewidth]{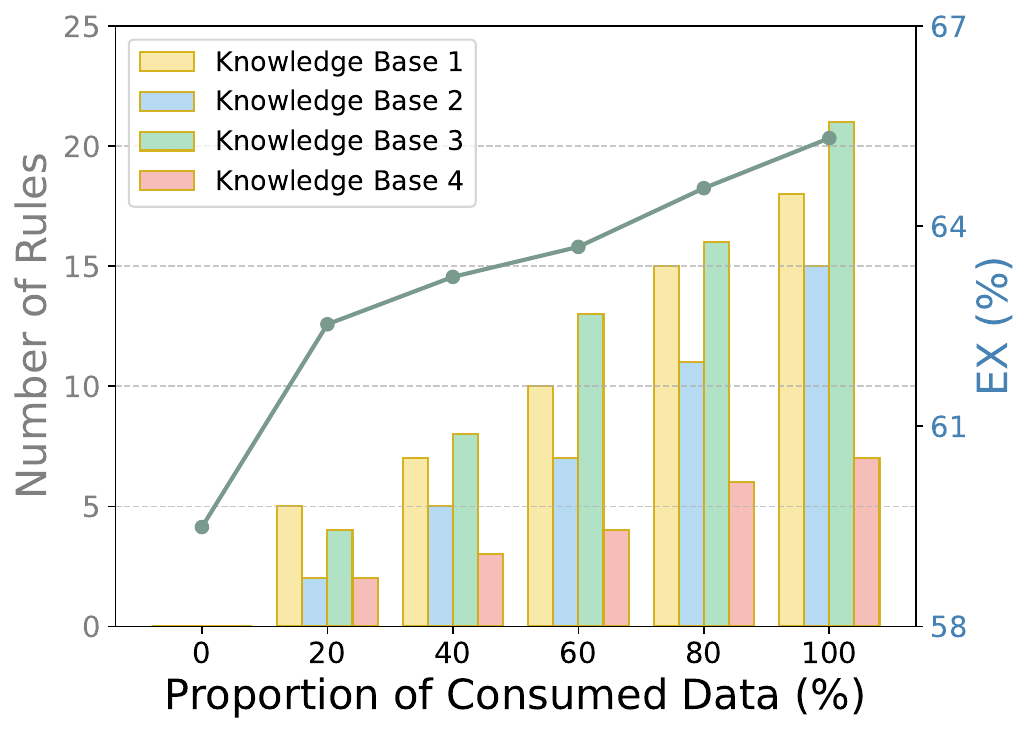}
        \caption{End-to-End}
        \label{bird_end2en}
    \end{subfigure}
    \vspace{-2.5mm}
    \caption{The performance metrics of \ours across different stages of self-learning. The bar chart corresponds to the left y-axis, while the line chart corresponds to the right y-axis.}
    \vspace{-6mm}
    \label{fig:percentage_increase}
\end{figure}

Specifically, for the end-to-end task, we use the predictive results of CodeS-15B on the BIRD's dev set.  
For SQL rewriting task, we choose the Payment-SQL dataset to examine the iterative gains of \ours in long SQL rewriting scenarios. 
The experimental results shown in Figure~\ref{fig:percentage_increase} demonstrate that the hybrid self-learning approach not only enhances the performance of \ours but also provides a reliable foundation for its continuous rewriting in real-world industrial applications. 
Moreover, the effectiveness of this mechanism further validates the feasibility of transitioning from expert-centric knowledge base construction to an expert-guided hybrid self-learning framework, thereby providing methodological support for reducing the cost of complete reliance on experts for knowledge collection and maintenance.

%% file: conclusions.tex
\section{Conclusions}

In this work, we present \ours, the first comprehensive LLM-based SQL toolkit with integrated knowledge management. It unifies four core functionalities—syntax correction, query rewriting, semantic refinement, and consistency verification—into a single framework powered by a hybrid self-learning mechanism.

One of the key innovations lies in its fragment-wise processing strategy. By focusing on individual fragments such as subqueries and CTEs, the approach improves precision while reducing the cognitive burden on LLMs.
Moreover, \ours incorporates an expert-guided hybrid self-learning framework that continuously enhances performance by extracting patterns from execution outputs and validating generated rules with minimal expert input.

Extensive experiments show that \ours consistently boosts base models' performance by up to 10\% in key metrics on benchmarks like BIRD and BIRD-CRITIC. Deployed in production environments, it demonstrates strong utility and adaptability across real-world databases.
